\documentclass[a4paper]{jpconf}
\usepackage[latin2]{inputenc}
\usepackage{t1enc}
\usepackage{graphicx}
\usepackage{cite}
\begin{document}
\title{Generalized DGLAP evolution}

\author{D Str\'ozik-Kotlorz$^{1,\, 2}$, S V Mikhailov$^2$ and O V Teryaev$^2$}

\address{$1$ Opole University of Technology, Division of Physics,
Pr\'oszkowska 76, 45-758 Opole, Poland}

\address{$2$ Bogoliubov Laboratory of Theoretical Physics, JINR,
         141980 Dubna, Russia}

\ead{dorota@theor.jinr.ru}

\begin{abstract}
We present progress in development of the truncated Mellin moments
approach (TMMA). We show our recent results on the generalization of DGLAP
evolution equations and discuss some their applications in spin physics.
\end{abstract}

\section{Introduction}
According to the factorization theorem, the cross sections for
DIS reactions and some classes of hadron - hadron collisions can be
expressed as convolution of two parts: a short-distance perturbative and
a long-distance nonperturbative ones. The perturbative part, describing
partonic cross sections at sufficiently high scale of the momentum transfer
$Q$ can be calculated within
perturbative chromodynamics (pQCD). The non-perturbative part contains
universal, process independent parton distribution functions  $f(x)$ (PDF) and
fragmentation functions  $D^h_q(x)$ (FF), which can be measured experimentally.
The evolution of these functions with the interaction scale $Q^2$ is again
described with the use of the perturbative QCD methods.
The standard DGLAP approach \cite{b1,b3,b4}
enables one to calculate parton densities which characterize the internal
nucleon structure at a given scale $Q^2$ when these densities are known
for a certain input scale $Q_0^2$.
We have shown that also the truncated Mellin moments
of the PDFs, $\int^1_z x^{n-1}f(x) dx$, satisfy the DGLAP evolution equations
and can be an additional tool in the QCD analysis of structure functions.
The major advantage of the TMMA is a possibility to adapt theoretical
QCD analysis to the experimentally available region of the Bjorken-$x$
variable. In this way, one can avoid the problem of dealing with the
unphysical region $x\rightarrow 0$ corresponding to the infinite energy
of interaction.
A number of important issues in particle physics, e.g., solving of the
`nucleon spin puzzle', quark - hadron duality or higher twist contributions
to the structure functions refers directly to moments.
Note that TMM, contrary to standard moments, may be directly extracted from the accurate
(JLab) data by appropriate binning (keeping $Q^2$ fixed).
These issues initiate a large number of experimental projects and
theoretical studies as well. Below we present the generalization of DGLAP
evolution equations within TMMA and discuss some applications in spin physics.

\section{Truncated Mellin moments approach}
The main finding of the TMMA is that the generalized truncated (cut) moments
(CMM) obtained by multiple integrations as well as
multiple differentiations of the original parton distribution also
satisfy the DGLAP equations with the simply transformed evolution kernel
\cite{b5,b6,b7}.
A similar generalized evolution equation, with the correspondingly modified
coefficient functions, can also be obtained for structure functions.
In Table~\ref{table 1}, we summarize the generalized CMM together with the
correspondingly transformed DGLAP evolution kernels.
\begin{table}
\caption{ \label{table 1}CMM (first column) and the corresponding evolution
kernels (second column).}
\begin{center}
\begin{tabular}{ll|l}
\br
  & Generalized CMM & DGLAP kernel ${\cal P}$    \\
\mr
1.&   $f(x)$                                      & $P(y)$           \\
  &                                               &                  \\
2.&   $x^nf(x)$                                   & $P(y)\cdot y^n$  \\
  &                                               &                  \\
3.&   $\int_{z}^{1}dx\,x^{n-1}\,f(x)$             & $P(y)\cdot y^n$  \\
  &                                               &                  \\
4.&   $\int\limits_z^1 z_{k}^{n_k-1} dz_k \: ...
\int\limits_{z_{2}}^{1} z_{1}^{n_1-1}\;f(z_1)\;dz_1$
& $P(y)\cdot y^{n_1+n_2+...+n_k}$  \\
  &                                               &                  \\
5.&   $f_\omega(z,n)= \left(\omega \ast fx^n\right)$
                         & $P(y)\cdot y^{n}$     \\
  &                                               &                  \\
6.&$\left( -\frac{d}{dx}\right)^k\left[x^n f(x)\right]$& $P(y)\cdot y^{n-k}$\\
\br
\end{tabular}
\end{center}
\end{table}
\vspace*{-5mm}

\section{Applications of TMMA}
Below we present examples of applications of TMMA to analysis of the
Bjorken sum rule \cite{b8} and fragmentation functions.

\subsection{Generalized Bjorken sum rule}
For any normalized function $\omega(x)$, $\int_{0}^{1}\omega(t)\, dt = 1$,
one can construct generalized CMM $f_\omega(z,n)$
as a Mellin convolution with the function $f$, which obeys the DGLAP
evolution equation with the rescaled kernel \cite{b9}:
\begin{equation}\label{eq.1}
f_\omega(z,n)= \left(\omega \ast fx^n\right)\equiv\int_{z}^{1}
\omega \left(z/x\right)~f(x)\,x^{n}\,dx/x\, ,
\end{equation}
%\begin{equation}\label{eq.2}
%\int_{0}^{1}\omega(t)\, dt = 1\, ,
%\end{equation}
\begin{equation}\label{eq.3}
{\cal P}(y) = P(y)\cdot y^{n}
\end{equation}
The special case of $f_\omega$ is suitable for the generalized Bjorken sum
rule (BSR):
\begin{equation}\label{eq.4}
G_\omega(x,Q^2)= \left(\omega \ast g_1^{NS}\right)(x).
\end{equation}
$G_\omega$ has the same evolution kernel as $g_1^{NS}$ and the generalized
BSR is equal to the ordinary BSR:
\begin{equation}\label{eq.5}
\int_{0}^{1}G_\omega(x,Q^2)\,dx = \int_{0}^{1} g_1^{NS}(x,Q^2)\, dx =
\rm{BSR}.
\end{equation}
The corresponding cut first moments of $G_\omega$ go to the BSR limit as the
cut point $x_0$ goes to zero.
This allows one to study behaviour of the generalized cut moments near
$x_0=0$ and estimate the value of the BSR from the cut integrals
$\int_{x_0}^1 G_\omega(x,Q^2)dx $ at $x_0 \neq 0$. The attempts for the case
$\omega(t)=n~t^{n-1}$ are shown in Fig.~1 and can be tested experimentally.
\begin{figure}[h]
\begin{center}
\begin{minipage}{15pc}
\includegraphics[width=15pc]{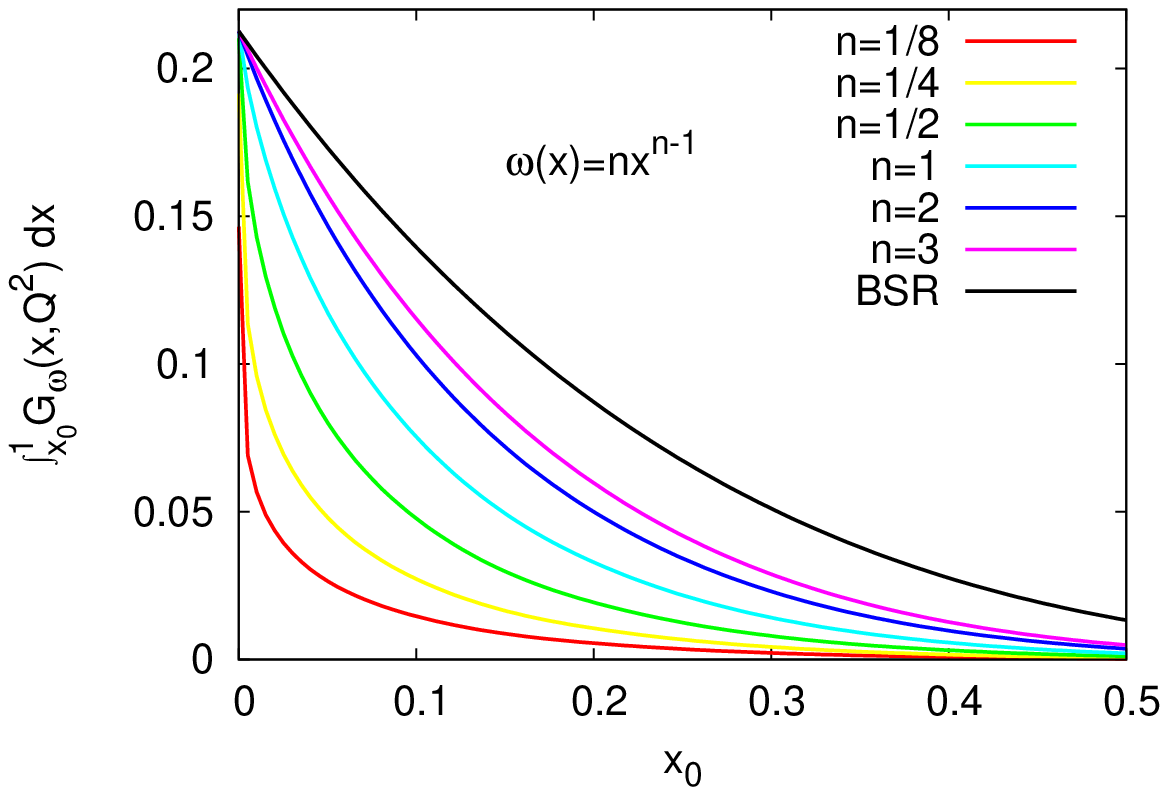}
\caption{\footnotesize \label{fig.1}The cut first moments of the generalized CMM
$G_n$ (\ref{eq.4}), where $\omega =n~t^{n-1}$, for different $n$ versus
the cut point $x_0$.}
\end{minipage}\hspace{1pc}%
\begin{minipage}{15pc}
\includegraphics[width=14pc]{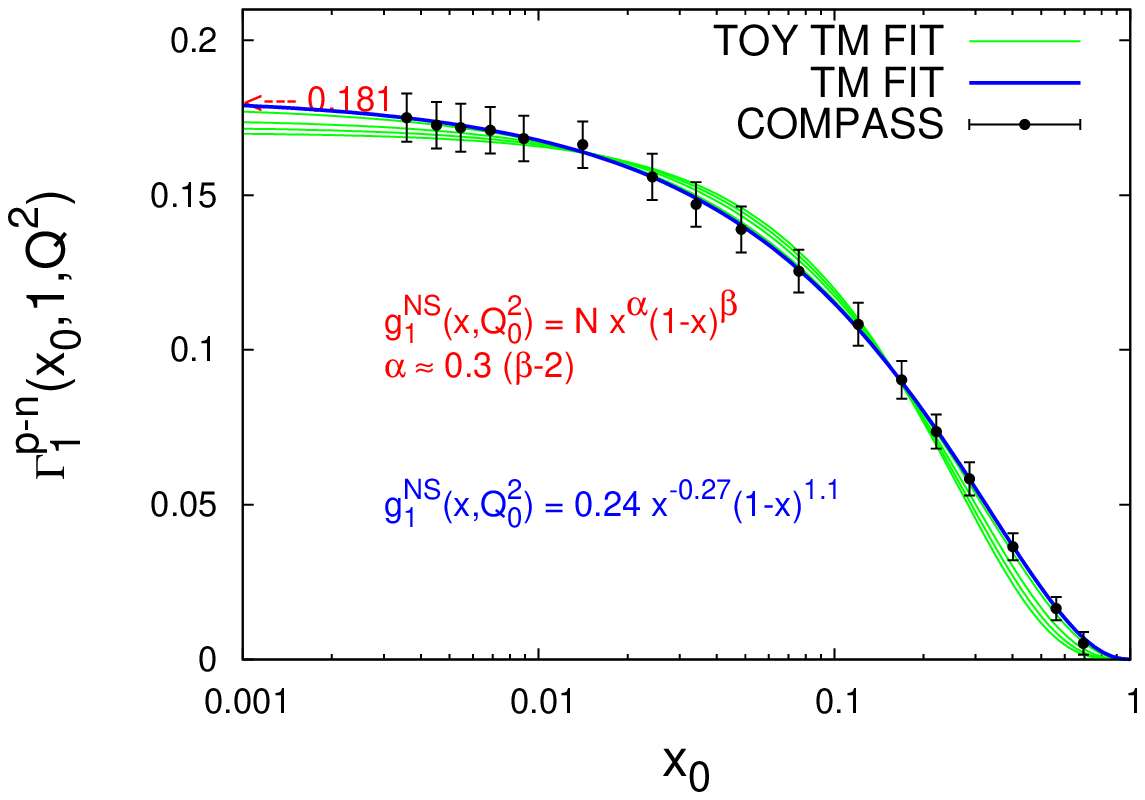}
\caption{\footnotesize \label{fig.2} Contributions to the Bjorken sum rule obtained within
TMMA. Comparison to COMPASS data.}
\end{minipage}
\end{center}
\end{figure}
We have also calculated contributions to the BSR itself and compared them to
the experimental data. In Fig.~2 we compare TMMA predictions for the
contributions to the BSR to recent COMPASS data. One can see that simple
input parametrization 7, where $\alpha = 0.3 (\beta -2)$ can
satisfactorily reproduce the experimental data. This relation between
$\alpha$ and $\beta$, together with the positivity constraint can provide
knowledge on the small-$x$ behaviour of the polarized structure function
$g_1^{NS}$. From our analysis the favoured small-$x$ behaviour of $g_1^{NS}$
is $x^\alpha$, where $\alpha = -0.2 \div -0.3$.
Table~\ref{table 2} contains the truncated contributions to the Bjorken sum
rule in the experimentally available $x$-region,
\begin{equation}\label{eq.6}
\Gamma_1^{p-n}(x_1,x_2,Q^2) = \int\limits_{x_1}^{x_2}g_1^{NS}(x,Q^2)\,dx
\end{equation}
obtained for different input parametrizations
\begin{equation}\label{eq.7}
g_1^{NS}(x,Q_0^2) = N\;x^{a_1}(1-x)^{a_2}.
\end{equation}
Our predictions are compared with the HERMES \cite{b10} and
COMPASS \cite{b11} data.
\begin{table}
\caption{\label{table 2} Truncated contributions to the Bjorken sum rule
with use different input parameterizations. Comparison with experimental
data.}
\begin{center}
\begin{tabular}{llll|ll}
\br
N&INPUT          & $x$ -range & $Q^2\,[\rm{GeV}^2]$ & $\Gamma_1^{NS}$ & EXP $\Gamma_1^{NS}$\\
\mr
1.&$(1-x)^3$ &                    &   & 0.161 & HERMES \\
2.&$x^{-0.2}(1-x)^3$ & 0.021-0.9  & $\:\:\:\:\:\:\:\:5$ & 0.149 & $0.1479\pm 0.0055\pm 0.0142$\\
3.&$x^{-0.4}(1-x)^3$ &            &   & 0.131 & \\
4.&$(1-x)^3$ &                    &   & 0.177 & COMPASS \\
5.&$x^{-0.2}(1-x)^3$ & 0.004-0.7  & $\:\:\:\:\:\:\:\:3$ & 0.173 & $0.175\pm 0.009\pm 0.015$\\
6.&$x^{-0.4}(1-x)^3$ &            &   & 0.163 & \\
7.&$x^{\alpha}(1-x)^{\beta}$& & & 0.168-& COMPASS \\
8.&$\alpha\approx 0.3(\beta-2)$ &
0.0025-0.7 & $\:\:\:\:\:\:\:\:3$ & 0.170 & $0.170\pm 0.008$ \\
\br
\end{tabular}
\end{center}
\end{table}
\vspace*{-5mm}
\subsection{Fragmentation functions}
Finally, it is worthy to mention that TMMA can be very useful in analysis
of the hadron fragmentation functions as in the small-$x$ region
behaviour of FF is known very poorly. In this way, one can restrict the
analysis to well determined models for $x\geq x_0$. FF also obey the
corresponding DGLAP evolution and their CMM can provide new insight into the hadron structure.
\begin{figure}[h]
\begin{center}
\begin{minipage}{16pc}
\includegraphics[width=14pc]{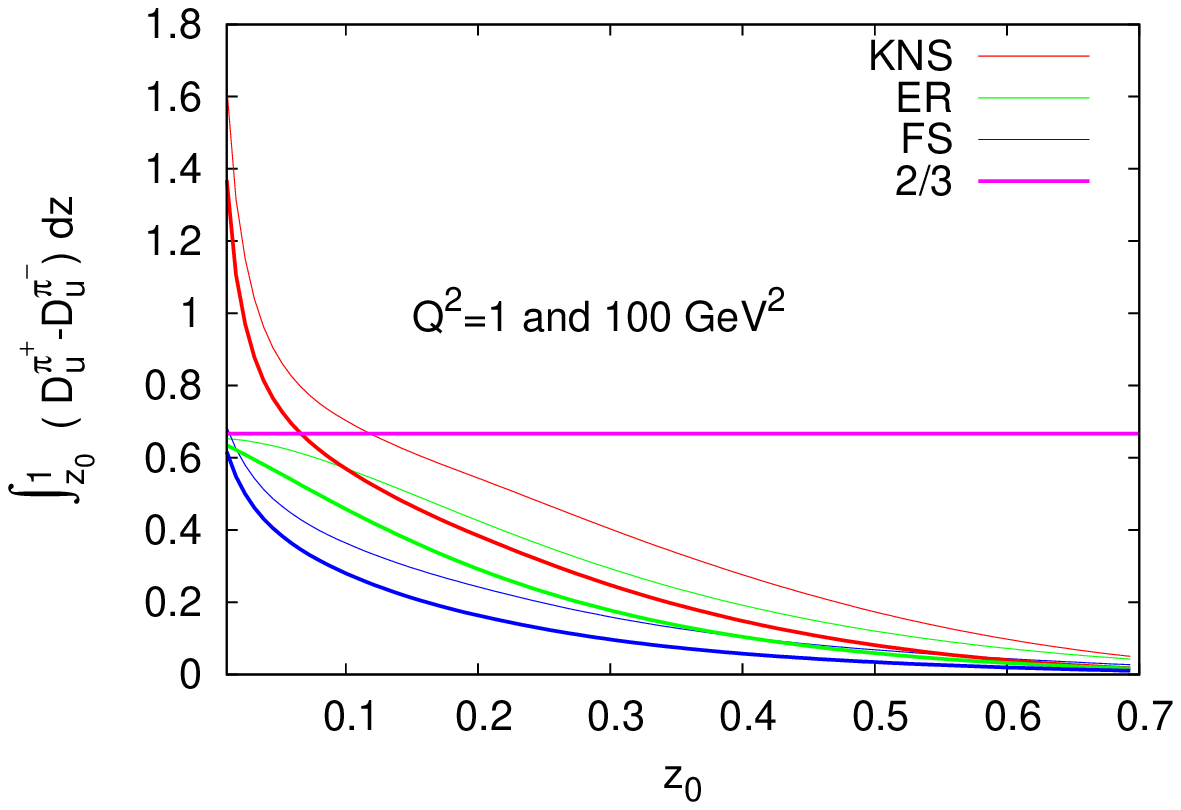}
\caption{\footnotesize \label{fig.3}TMM of the pion~FF, $Q^2=1,\, 100\; \rm{GeV}^2$ (thin,
thick). Inputs: KNS~\cite{b12}, ER~\cite{b13}, FS~\cite{b14}}
\end{minipage}\hspace{1pc}%
\begin{minipage}{16pc}
\includegraphics[width=14pc]{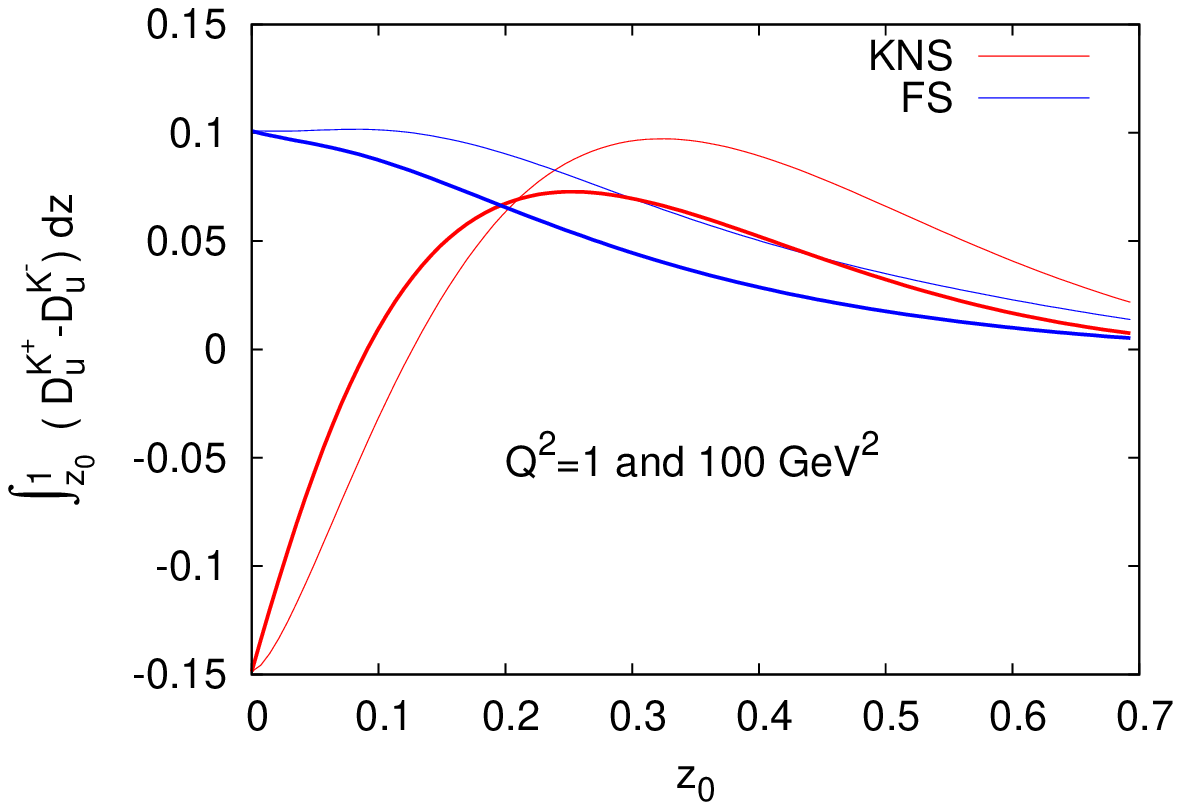}
\caption{\footnotesize \label{fig.4}TMM of the kaon~FF, $Q^2=1,\, 100\; \rm{GeV}^2$ (thin,
thick). Inputs: KNS~\cite{b12}, FS~\cite{b14}}
\end{minipage}
\end{center}
\end{figure}
\newline
In Figs.~3,4 we present evolution of the truncated moments of FF contributing
to the quark charge conservation,
\begin{equation}
\label{eq.8}
\sum_h Q_h \int_0^1\; dz   D_q^h (z,Q^2)   = Q_q ,
\end{equation}
where $D_q^h$ denotes a fragmentation function of the hadron $h$ from a
parton $q$ and $Q_h, Q_q$ are the charges of the hadron $h$ and parton $q$.
%By virtue of the charge symmetry, one has
%\begin{equation}\label{eq.9}
%D_{u}^{\pi^+,K^+} (z,Q^2) = D_d^{\pi^-,K^-} (z,Q^2)\, .
%\end{equation}
We use different parametrization of the FF at the initial scale \cite{b12,b13,b14}.
The excess of the obtained moments for pions, providing the main contributions to sum rule, over the charge conservation values at
$z_0 \sim 0.2$ may be considered as a support of inapplicability of independent fragmentation
picture at this region.
This fact is also exhibited in the large differences in the presented
predictions in this region, depending on the input
parametrization, for both pions and kaons.
%This is an evidence of a lack of knowledge on the small-$z_0$
%behaviour of FF.
Therefore TMM approach can be a natural tool also in the
study of the fragmentation functions, requiring further investigations.

\ack
We are grateful to A.V. Efremov to stimulating discussions. O.T. is indebted to S. Kumano for useful discussions.
This work is supported by the Bogoliubov-Infeld Program,
Grant No. 01-3-1113-2014/2018 and by the Russian Foundation for Fundamental
Research, Grant No. 14-01-00647a.

\end{document}